\documentclass[aps,preprint,showpacs,preprintnumbers,showkeys,eqsecnum,amsmath,amssymb]{revtex4}

\usepackage{txfonts}
\usepackage{dcolumn}
\usepackage{mathrsfs}
\usepackage{bm}
\usepackage{amsmath,amssymb,epsfig,float}

\begin{document}

\title{How the Hawking effect and prepared states affect Entanglement distillability of Dirac fields}

\author{Junfeng Deng, Jieci Wang and Jiliang Jing\footnote{Corresponding author, Email: jljing@hunnu.edu.cn}}
\affiliation{ Institute of Physics and  Department of Physics,
\\ Hunan Normal University, Changsha, \\ Hunan 410081, P. R. China
\\ and
\\ Key Laboratory of Low-dimensional Quantum Structures
\\ and Quantum
Control of Ministry of Education, \\ Hunan Normal University,
Changsha, Hunan 410081, P. R. China}

\vspace*{0.2cm}
\begin{abstract}
\vspace*{0.2cm}

How the Hawking effect and the prepared states influence the entanglement distillability of Dirac fields in the Schwarzschild spacetime is studied by using the Werner states which are composed of the maximum or generic entangled states. It is found that the states are entangled when the parameter of the Werner states, $F$, satisfies  $\tau<F\leq 1$ in which $\tau$ is influenced both by the Hawking temperature of the black hole and energy of the fields. It is also shown that although the parameter of the generic entangled states, $\alpha$, affects the entanglement, it does not change the range of the parameter, $F$, where the states are entangled  for the case of generic entangled states.

\end{abstract}

\vspace*{1.5cm}


\maketitle

\section{INTRODUCTION}

Recently, many people have paid their attentions to the
investigation of the effects of relativity on quantum information
\cite{Peres,Ge-Kim,I. Fuentes-Schuller,adesso,adesso2,P. M. Alsing,Alsing-McMahon-Milburn,I.
Lamata,Pan, J.L,Jieci, A. D,Jieci W, Hawking-Terashima, Shapoor Moradi, Qiyuan Pang, Jieci Wang, E.M} because it is important to both the theoretical physics and practical application.
The behaviors of entanglement in the noninertial frame were discussed
\cite{Peres,Ge-Kim,P. M. Alsing,Alsing-McMahon-Milburn,I. Fuentes-Schuller, adesso, adesso2, I. Lamata,Pan,J.L,Jieci W,Hawking-Terashima}, and the these studies showed that the entanglement of the states will be degraded for an accelerated observer because the Unruh effect \cite{W. G. Unruh} results in a loss of the information. More recently, in order to further investigate characterization of a quantum system Datta \cite{A. D} and Wang {\it et al}\cite{Jieci} calculated the quantum discord
between two relatively accelerated scalar and Dirac modes respectively. Shahpoor Moradi \cite{Shapoor Moradi} studied the entanglement distillability of a bipartite mixed states as seen by two relatively accelerated
parties. The quantum information in the background of black
holes was investigated \cite{Qiyuan Pang,Jieci Wang,E.M}. It was found that the entanglement and fidelity of the teleportation are
degraded by the Hawking effect \cite{S. W. Hawking}. Eduardo Mart\'{i}n-Mart\'{i}nez {\it et al}
\cite{E.M} studied the behavior of quantum entanglement near a black hole, and they showed the entanglement is  a function of the mass of the black hole, the frequency of the field and the physical distance from the noninertial observer to the horizon. M. Aspachs {\it et al} \cite{Aspachs} analysed the general quantum-statistical grounds the problem of optimal detection of the Unruh-Hawking effect. Downes {\it et al} \cite{Downes} investigated the entangling moving cavities in non-inertial frames. And Martin-Martinez {\it et al} \cite{Martin-Martinez} analysed the quantum entanglement produced in the formation of a black hole.

In this paper, we study the entanglement distillability of the Dirac fields in the Schwarzschild spacetime. We focus our attention on how the Hawking effect and prepared states influence the entanglement  distillability and separability. In order to illustrate this problem convenient, we take the Werner states which are composed with the maximum or generically entangled states. It should point out here that there are two state parameters ($F$, $\alpha$) and different value for these parameters represents different prepared states. We also assume that two
observers, Alice and Rob, share the Werner states in flat Minkowski
spacetime.  Alice  moves along a geodesic with a detector which only detects mode $k$, while Rob hovers outside the event horizon with a
uniform acceleration with a detector sensitive only to mode $s$.
With the result of Rob's detects, we want to see whether the
characteristic of the entanglement distillability changes.

The outline of the paper is as follows. In Sec. II we discuss the
relationship between kruskal vacuum and Schwarzschild vacuum. In
Sec. III we investigate the entanglement distillability of the
Werner states composed with maximum entangled states. In Sec. IV we
discuss the entanglement distillability of the Werner states
composed with generically entangled states. In the last section, we
summarize and discuss our conclusions.

\section{RELATIONSHIP BETWEEN KRUSKAL VACUUM AND SCHWARZSCHILD VACUUM}

The Dirac equation in curve spacetime can be written as \cite{D.
R. Brill}
\begin{eqnarray} \label{Dirac}
[\gamma^{a}e_a^\mu(\partial_\mu+\Gamma_\mu)]\Psi=0,
\end{eqnarray}
here $\gamma^{a}$ is the Dirac matric,  $e_a^\mu$ represents the
inverse of the tetrad $e^a_\mu$, and
$\Gamma_\mu=\frac{1}{8}[\gamma^a,\gamma^b]e_a^{\nu}e_ {b\nu;\mu}$ is
the spin connection coefficient. Hereafter we use
$G=c=h=\kappa_B=1$.

We solve Eq. (\ref{Dirac}) in the Schwarzschild spacetime and get
the positive frequency outgoing solutions $
\Psi^{I+}_{k_i}\sim e^{-i \omega_i u}$ and
$\Psi^{II+}_{k_i}\sim e^{i \omega_i u}$, $\forall~ i$ \cite{Jieci Wang}, where
$u=t-r-2 M \ln[(r-2 M)/(2 M)]$ and $\omega_i$ is a monochromatic frequency of the
Dirac field. Since $\Psi^{I+}_{k_i}$ and
$\Psi^{II+}_{k_i}$ are analytic outside and inside the event
horizon respectively, they form a complete orthogonal family. Thus,
we can expand the Dirac field $\Psi_{out}$ as
\begin{eqnarray}\label{First expand}
&&\Psi_{out}=\sum_{i, \sigma}\int
dk[a^{(\sigma)}_{k_i}\Psi^{(\sigma)+}_{k_i}
+b^{(\sigma)\dag}_{k_i}\Psi^{(\sigma)-}_{k_i}],
\end{eqnarray}
where $\sigma=(I,II)$,  $a^{I}_{k_i}$ and
$b^{I\dag}_{k_i}$ are the fermion annihilation and
antifermion creation operators acting on the state of the exterior
region, and $a^{II}_{k_i}$ and $b^{II\dag}_{k_i}$ are
the fermion annihilation and antifermion creation operators acting
on the state of the interior region, respectively.

On the other hand, introducing Kruskal coordinates $\mathcal {U}$
and $\mathcal {V}$ for the Schwarzschild spacetime
\begin{eqnarray}
u&=&-4M\ln(-\frac{\mathcal {U}}{4M}),~~~ v=4M\ln(\frac{\mathcal
{V}}{4M}),\ \ {\text  if}\ \ r>r_+, \nonumber
\\
u&=&-4M\ln(\frac{\mathcal {U}}{4M}),~~~~~ v=4M\ln(\frac{\mathcal
{V}}{4M}),\ \ {\text if}\ \ r<r_+,
\end{eqnarray}
and making an analytic continuation for $\Psi^{I+}_{k_i}$
and $\Psi^{II+}_{k_i}$, we find a complete basis for positive
energy modes  which analytic for all real $\mathcal {U}$ and
$\mathcal {V}$ according to the suggestion of Damour-Ruffini

\begin{eqnarray}
&&\mathscr{F}^{I+}_{k_i}=e^{2\pi
M\omega_i}\Psi^{I+}_{k_i} +e^{-2\pi
M\omega_i}\Psi^{II-}_{-k_i}, \nonumber
\\
&&\mathscr{F}^{II+}_{k_i}=e^{-2\pi
M\omega_i}\Psi^{I-}_{-k_i} +e^{2\pi
M\omega_i}\Psi^{II+}_{k_i}.
\end{eqnarray}
Thus, we can also expand the Dirac fields in the Kruskal spacetime
as
\begin{eqnarray}\label{Second expand}
\Psi_{out}=\sum_{i,\sigma}\int dk[2\cosh(4\pi
M\omega_i)]^{-1/2}[c^{(\sigma)}_{k_i}
\mathscr{F}^{(\sigma)+}_{k_i}+d^{(\sigma)\dag}_{k_i}\mathscr{F}^{(\sigma)-}
_{k_i} ],
\end{eqnarray}
where $c^{\sigma}_{k_i}$ and $d^{\sigma\dag}_{k_i}$
are the annihilation and creation operators acting on the Kruskal
vacuum. It is worth mentioning that there exist an infinite number of combination of Kruskal annihilation operators $c^I_{k_j}=\sum_i C_{ij}\, c^I_{k_i}$ (where $C_{ij}$ are coefficients) annihilates the same Kruskal vacuum state $|0\rangle_{\mathcal{K}}$ [18,19].

Eqs. (\ref{First expand}) and (\ref{Second expand}) represent the
decomposition of the Dirac fields in Schwarzschild  and Kruskal
modes respectively, so we can get the Bogoliubov transformations
\cite{Barnett} between the Schwarzschild and Kruskal creation and
annihilation operators. In consideration of the Bogoliubov
relationships being diagonal, each annihilation operator $c^I_{k_i}$ can be expressed as a combination of Schwarzschild particle operators of only one
Schwarzschild frequency $\omega_i$ \cite{E.M, Bruschi}
\begin{eqnarray}\label{Bog1}
c^I_{k_i}=(e^{-8\pi M\omega_i}+1)^{-\frac{1}{2}} a^I_{k_i}-(e^{8\pi
M\omega_i}+1)^{-\frac{1}{2}}b^{II\dag} _{k_i}.
\end{eqnarray}
Then the Kruskal vacuum and  excited states can express by
Schwarzschild Fock space basis
$|0\rangle_{\mathcal {K}}=\bigotimes_i|0_{k_i}\rangle_{\mathcal {K}}
$ and $|1\rangle_{\mathcal {K}}=\bigotimes_i|1_{k_i}\rangle_{\mathcal {K}}$, where
\begin{eqnarray}\label{p0}
&&|0_{k_i}\rangle_{\mathcal {K}}=
(e^{-\omega_i/T}+1)^{-\frac{1}{2}}|0_{k_i}
\rangle_{I}|0_{-k_i}\rangle_{II}+(e^{\omega_i/T}+1)^{-\frac{1}{2}}|1_{k_i}
\rangle_{I}|1_{-k_i}\rangle_{II}, \\ \label{p1}
&&|1_{k_i}\rangle_{\mathcal{K}}=c^{I\dag}_{k_i}|0_{k_i}\rangle_{\mathcal
{K}}=|1_{k_i}\rangle_{I}|0_{-k_i}\rangle_{II},
\end{eqnarray}
where the $|n_{k_i}\rangle_{I}$ and $|n_{-k_i}\rangle_{II}$
are the orthogonal bases for the inside and outside region of the event horizon respectively, and $T=\frac{1}{8\pi M}$ is the Hawking temperature \cite{R.
Kerner}.

\section{ENTANGLEMENT distillability FOR WERNER STATES
COMPOSED WITH MAXIMUM ENTANGLED STATES}

To study the entanglement distillability in the
Schwarzschild spacetime, we consider a Dirac field which is, from an inertial perspective, in a special superposition of Kruskal monochromatic modes (see \cite{E.M, Bruschi}  for details) such that, in the Schwarzschild frames, the observers detect the field in the single-mode states. The two
modes share the Werner states which can be written as \cite{R. F.
Werner}
\begin{eqnarray} \label{ARR}
\rho_{AR}=F|\psi^-\rangle_{AR}\langle \psi^-|+\frac{1-F}{3}
(|\phi^+\rangle_{AR}\langle\phi^+|+|\phi^-\rangle_{AR}
\langle\phi^-|+ |\psi^+\rangle_{AR}\langle\psi^+|),
\end{eqnarray}
where F is a parameter (runs from $0$ to $1$) which gives different prepared states, and $|\phi^\pm\rangle$ and $|\psi^\pm\rangle$ are usual entangled Bell states
\begin{eqnarray}
&&|\phi^{\pm}\rangle_{AR}=\frac{1}{\sqrt{2}}(|0\rangle_A|0\rangle_R
\pm|1\rangle_A|1_{k_R}\rangle_R), \nonumber
\\ &&
|\psi^{\pm}\rangle_{AR}=\frac{1}{\sqrt{2}}(|0
\rangle_A|1_{k_R}\rangle_R\pm|1\rangle_A|0\rangle_R),
\end{eqnarray}
where $|0\rangle_A(R)$ is vacuum states from a inertial viewpoint,
$|1_{k_R}\rangle_R$ is a single particle excitation state which characterized by the frequency $\omega_R$ observed by Rob, and all other modes of the field are
in vacuum state. Hereafter we will refer to the frequency $\omega_R$ simply as
$\omega$.

We can see from Eq. (\ref{ARR}) that the Werner states are
characterized by the parameter $F$. In a inertial frame, it was
shown that for $F\leq{\frac{1}{2}}$ the Werner states are
separable, while for $\frac{1}{2}\leq{F}\leq{1}$ are entangled
\cite{R. F. Werner}.

Because Alice is an inertial observer and Rob hovers over the event
horizon with a constant acceleration, the states corresponding to
modes $s$ must be expanded by the bases for the inside and outside
region of the event horizon in
order to describe what Rob see. From the discussion in the section
$II$, we know that the Kruskal states $|0\rangle_{\mathcal{K}}$ and
$|1\rangle_{\mathcal{K}}$ are correspond to two mode states in the
Schwarzschild frame described by Eqs. (\ref{p0}) and (\ref{p1}),
then the density matrix $\rho_{AR}$ takes the form
\begin{eqnarray}
\rho_{A,I,II}=  \left(
          \begin{array}{cccccccc}
            \frac{1-F}{3(e^{\frac{-\omega}{T}}+1)} & 0 & 0 & \frac{(1-F)(e^{\frac{-\omega}{T}}+1)^{\frac{-1}{2}}}{3(e^{
            \frac{\omega}{T}}+1)^{\frac{1}{2}}} & 0 & 0 & 0 & 0 \\
            0 & 0 & 0 & 0 & 0 & 0 & 0 & 0\\
            0 & 0 & \frac{2F+1}{6} & 0 & \frac{(1-4F)}{6(e^{\frac{-\omega}{T}}+1)^{\frac{1}{2}}} & 0 & 0 & \frac{1-4F}{6(e^{\frac{\omega}{T}}+1)^{\frac{1}{2}}} \\
            \frac{(1-F)(e^{\frac{-\omega}{T}}+1)^{\frac{-1}{2}}}{3(e^{
            \frac{\omega}{T}}+1)^{\frac{1}{2}}} & 0 & 0 & \frac{1-F}{3(e^{\frac{\omega}{T}}+1)} & 0 & 0 & 0 & 0 \\
            0 & 0 & \frac{1-4F}{6(e^{\frac{-\omega}{T}}+1)^{\frac{1}{2}}} & 0 & \frac{1+2F}{6(e^{\frac{-\omega}{T}}+1)} & 0 & 0 &
             \frac{(1+2F)(e^{\frac{-\omega}{T}}+1)^{\frac{-1}{2}}}{6(e^{
            \frac{\omega}{T}}+1)^{\frac{1}{2}}} \\
            0 & 0 & 0 & 0 & 0 & 0 & 0 & 0 \\
            0 & 0 & 0 & 0 & 0 & 0 & \frac{1-F}{3} & 0 \\
            0 & 0 & \frac{1-4F}{6(e^{\frac{\omega}{T}}+1)^{\frac{1}{2}}} & 0 &
           \frac{(1+2F)(e^{\frac{-\omega}{T}}+1)^{\frac{-1}{2}}}{6(e^{
            \frac{\omega}{T}}+1)^{\frac{1}{2}}} & 0 & 0 & \frac{1+2F}{6(e^{\frac{\omega}{T}}+1)} \\
          \end{array}
        \right),
\end{eqnarray}
where the matrix is written on the basis of $|0\rangle_A|0\rangle_I
|0\rangle_{II}$, $|0\rangle_A|0\rangle_I
|1\rangle_{II}$, $|0\rangle_A|1\rangle_I
|0\rangle_{II}$, $|0\rangle_A|1\rangle_I
|1\rangle_{II}$, $|1\rangle_A|0\rangle_I
|0\rangle_{II}$, $|1\rangle_A|0\rangle_I
|1\rangle_{II}$, $|1\rangle_A|1\rangle_I
|0\rangle_{II}$, $|1\rangle_A|1\rangle_I
|1\rangle_{II}$. We have to trace over the states in this region because Rob is
causally disconnected from the interior  region of the black hole.
Thus, we get the mixed density matrix between Alice and Rob
\begin{eqnarray}
\rho_{A,I}=\left(
  \begin{array}{cccc}
    \frac{1-F}{3(e^{\frac{-\omega}{T}}+1)} & 0 & 0 & 0 \\
    0 & \frac{1+2F}{6}+\frac{1-F}{3(e^{\frac{\omega}{T}}+1)} & \frac{(1-4F)}{6(e^{\frac{-\omega}{T}}+1)^{\frac{1}{2}}} & 0 \\
    0 & \frac{(1-4F)}{6(e^{\frac{-\omega}{T}}+1)^{\frac{1}{2}}} & \frac{1+2F}{6(e^{\frac{-\omega}{T}}+1)} & 0 \\
    0 & 0 & 0 & \frac{1-F}{3}+\frac{1+2F}
    {6(e^{\frac{\omega}{T}}+1)}
  \end{array}
\right),
\end{eqnarray}
where the matrix is written on the basis of $|0\rangle_A|0\rangle_I
$, $|0\rangle_A|1\rangle_I
$, $|1\rangle_A|0\rangle_I
$, $|1\rangle_A|1\rangle_I
$. To determine whether the mixed states are the entanglement states or
not, we use the partial transpose criterion \cite{A. Peres} which
states that the states are entanglement one if one eigenvalue of the
partial transpose is negative at least. It worth to mention  that
the states with positive partial transpose density matrix  have no
distillable entanglement, but may have nondistillable entanglement if
the dimension is larger than $2$. We find the eigenvalues of the partial
transpose of $\rho_{AR}$ are
\begin{eqnarray}
\lambda^{(1)}&=&\frac{1+2F}{6(e^{\frac{-\omega}{T}}+1)},
\\
\lambda^{(2)}&=&\frac{1+2F}{6}+\frac{1-F}{3(e^{\frac{\omega}{T}}+1)},
\\
\lambda^{(3)}&=&\frac{1-F}{3}+\frac{4F-1}{12}\bigg(\frac{1}
{e^{\frac{\omega}{T}}+1}+2\sqrt{\frac{1}{e^{\frac{-\omega}
{T}}+1}+\frac{3}{2(e^{\frac{\omega}{T}}+1)(4F-1)}}\bigg),
\\
\lambda^{(4)}&=&\frac{1-F}{3}+\frac{4F-1}{12}\bigg(\frac{1}
{e^{\frac{\omega}{T}}+1}-2\sqrt{\frac{1}{e^{\frac{-\omega}{T}}
+1}+\frac{3}{2(e^{\frac{\omega}{T}}+1)(4F-1)}}\bigg).
\end{eqnarray}
\begin{figure}[ht]
\includegraphics[scale=0.82]{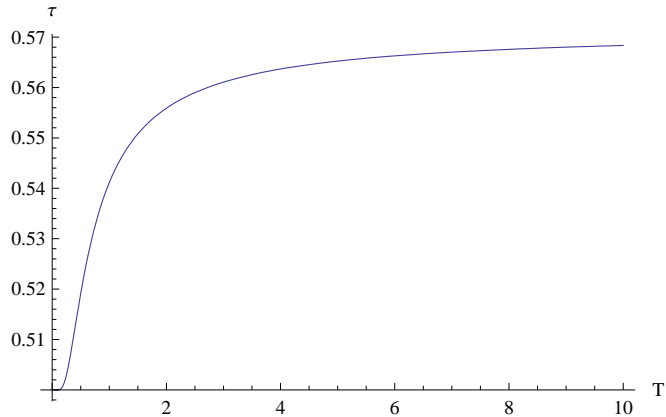}
\caption{\label{entropy}Plot of $\tau=\frac{3e^{\omega/T}+5}
{6e^{\omega/T}+8}$ as a function of the Hawking temperature (we take $\omega=1$.)}
\end{figure}
It is clearly that $\lambda^{(1)}$ $\lambda^{(2)}$ and
$\lambda^{(3)}$ are positive but  $\lambda^{(4)}$ is negative when
$F>\tau$ with $\tau=\frac{3e^{\omega/T}+5} {6e^{\omega/T}+8}$.

From these eigenvalues we know that the states are entangled for
$\tau<F\leq 1$ in the Schwarzschild spacetime, which is different
from that $1/2<F\leq 1$ in the inertial frame. It is obvious that
the Hawking temperature $T$ and energy $\omega$ of the states play
an important role in the low boundary $\tau$.

In Fig. 1 we study how the Hawking temperature $T$  changes $\tau$,
and see that $\tau$ increases with the increases of the temperature $T$.
If the Hawking temperature becomes zero, $T=0$, we have $\tau=1/2$.
Thus, for a system with zero temperature the states are entangled
for $1/2<F\leq1$, which is the same as that in the inertial frame.
But in the limit of $T\rightarrow\infty,$ we get $\tau=0.57$. That
is to say, when $T\rightarrow\infty$ the states are entangled for $0.57<F\leq1$.

In Fig. 2 we plot $\tau$ as a function of $\omega$. We learn from
the figure that, unlike in inertial frame, the energy of the states
affects the distillability of the entanglement. It is shown that
$\tau$  decrease with the increase of $\omega$. In particular,
$\tau=0.57$ as $\omega\rightarrow 0$ and $\tau=1/2$ as
$\omega\rightarrow \infty$.

\begin{figure}[ht]
\includegraphics[scale=0.82]{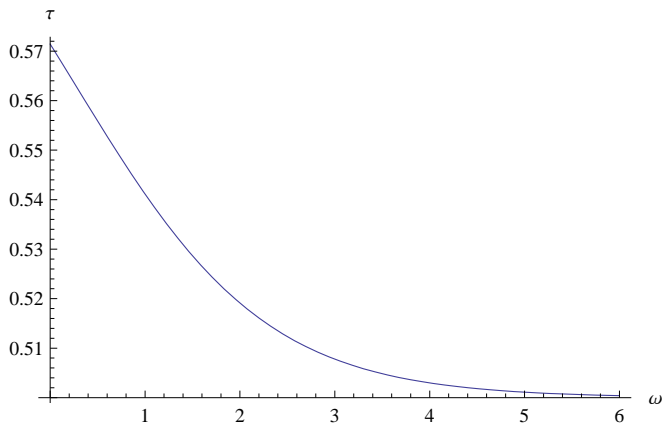}
\caption{\label{entropy1}Plot of $\tau=\frac{3e^{\omega/T}+5}
{6e^{\omega/T}+8}$ as a function of the energy  $\omega$
(we take $T=1$.)}
\end{figure}

\section{ENTANGLEMENT DISTILLABILITY FOR WERNER STATES COMPOSED
 WITH generically ENTANGLED STATES}
We now consider the Werner states which are composed with generically
entangled states
\begin{eqnarray}\label{ARU}
&&|\phi^{\pm}\rangle_{AR}=\alpha|0\rangle_A|0\rangle_R
\pm\sqrt{1-\alpha^2}|1\rangle_A|1_{k_R}\rangle_R, \nonumber
\\ &&
|\psi^{\pm}\rangle_{AR}=\alpha|0
\rangle_A|1_{k_R}\rangle_R\pm\sqrt{1-\alpha^2}|1\rangle_A|0\rangle_R),
\end{eqnarray}
where $\alpha$ is a real number that satisfies $0<\alpha<1$, and
$\alpha$ and $\sqrt{1-\alpha^2}$ are the so-called normalized
partners.

The main purpose of this section is to see whether the parameter
$\alpha$ influences the entanglement distillability. Using Eqs.
(\ref{p0}) and (\ref{p1}) we obtain the following density matrix
$\rho_{AR}$
\begin{eqnarray}
\rho_{A,I,II} =      \left(
          \begin{array}{cccccccc}
            \frac{2\alpha^2(1-F)}{3(e^{\frac{-\omega}{T}}+1)} & 0 & 0 & \frac{2\alpha^2(1-F)A}{3} & 0 & 0 & 0 & 0 \\
            0 & 0 & 0 & 0 & 0 & 0 & 0 & 0\\
            0 & 0 & \frac{\alpha^2(2F+1)}{3} & 0 & \frac{\alpha\sqrt{1-\alpha^2}(1-4F)}{3(e^{\frac{-\omega}{T}}+1)^{\frac{1}{2}}} & 0 & 0 & \frac{\alpha\sqrt{1-\alpha^2}(1-4F)}{3(e^{\frac{\omega}{T}}+1)^{\frac{1}{2}}} \\
            \frac{2\alpha^2(1-F)A}{3
            } & 0 & 0 & \frac{\alpha^2(2-2F)}{3(e^{\frac{\omega}{T}}+1)} & 0 & 0 & 0 & 0 \\
            0 & 0 & \frac{\alpha\sqrt{1-\alpha^2}(1-4F)}{3(e^{\frac{-\omega}{T}}+1)^{\frac{1}{2}}} & 0 & \frac{(1-\alpha^2)(1+2F)}{3(e^{\frac{-\omega}{T}}+1)} & 0 & 0 &
             \frac{(1-\alpha^2)(1+2F)A}{3
            } \\
            0 & 0 & 0 & 0 & 0 & 0 & 0 & 0 \\
            0 & 0 & 0 & 0 & 0 & 0 & \frac{2(1-\alpha^2)(1-F)}{3} & 0 \\
            0 & 0 & \frac{\alpha\sqrt{1-\alpha^2}(1-4F)}{3(e^{\frac{\omega}{T}}+1)^{\frac{1}{2}}} & 0 &
           \frac{(1-\alpha^2)(1+2F)A}{3
            } & 0 & 0 & \frac{(1-\alpha^2)(1+2F)}{3(e^{\frac{\omega}{T}}+1)} \\
          \end{array}
        \right),
\end{eqnarray}
where
$A=\frac{(e^{\frac{-\omega}{T}}+1)^{\frac{-1}{2}}}{(e^{\frac{\omega}{T}}+1)^{\frac{1}{2}}}$.
Tracing over the modes in the region II we find the density matrix
between Alice and Rob
\begin{eqnarray}
\rho_{A,I}=\left(
  \begin{array}{cccc}
    \frac{2\alpha^2(1-F)}{3(e^{\frac{-\omega}{T}}+1)} & 0 & 0 & 0 \\
    0 & \frac{(1+2F)\alpha^2}{3}+\frac{2\alpha^2(1-F)}{3(e^{\frac{\omega}{T}}+1)} & \frac{\alpha\sqrt{1-\alpha^2}(1-4F)}{3(e^{\frac{-\omega}{T}}+1)^{\frac{1}{2}}} & 0 \\
    0 & \frac{\alpha\sqrt{1-\alpha^2}(1-4F)}{3(e^{\frac{-\omega}{T}}+1)^{\frac{1}{2}}} & \frac{(1-\alpha^2)(1+2F)}{3(e^{\frac{-\omega}{T}}+1)} & 0 \\
    0 & 0 & 0 & \frac{2(1-\alpha^2)(1-F)}{3}+\frac{(1-\alpha^2)(1+2F)}
    {3(e^{\frac{\omega}{T}}+1)}
  \end{array}
\right).
\end{eqnarray}

To show how $\alpha$ affects the entanglement, we use the
logarithmic negativity which is defined as
$N(\rho_{A,I})=log_2\|\rho_{A,I}^T\|$. $N(\rho_{A,I})$ as a function of the parameters
$F$ and $\alpha$ is plotted in Fig. 3.
\begin{figure}[ht]
\includegraphics[scale=0.82]{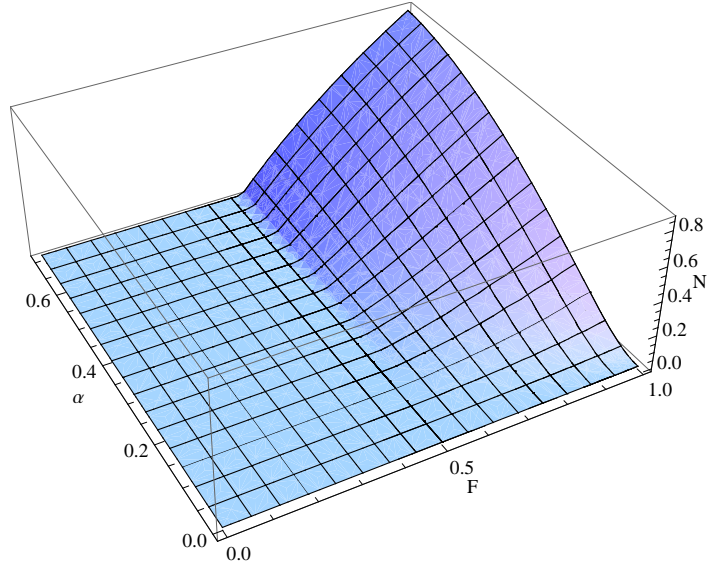}
\caption{\label{Logrithmic}Plot of logarithmic negativity $N$
 as a function of the parameter $F$ and $\alpha$ with $\omega/T=3/2$.}
\end{figure}

The eigenvalues of the partial transpose of the $\rho_{A,I}$ are
\begin{eqnarray}
\lambda^{(1)}&=&\frac{(1+2F)\alpha^2}{3}+\frac{2\alpha^2(1-F)}
{3(e^{\frac{\omega}{T}}+1)},\\
\lambda^{(2)}&=&\frac{(1-\alpha^2)(1+2F)}
{3(e^{\frac{-\omega}{T}}+1)},\\
\lambda^{(3)}&=&\frac{1}{6}[3-3\alpha^2+B^2(-2F-1+3\alpha^2)
+\sqrt{\xi}],\\
\lambda^{(4)}&=&\frac{1}{6}[3-3\alpha^2+B^2(-2F-1+3\alpha^2)
-\sqrt{\xi}],
\end{eqnarray}
where
$\xi=9(\alpha^2-1)^2-2B^2(\alpha^2-1)[(4F-1)(8F+1)\alpha^2-6F-3]+B^4[1+\alpha^2+F(2-4\alpha^2)]^2$
and $ B=(e^{\frac{-\omega}{T}}+1)^{\frac{-1}{2}}$. The eigenvalue
$\lambda^{(4)}$ is negative when $F>\sigma$ with
\begin{eqnarray}
\sigma=\frac{3e^{\omega/T}+5} {6e^{\omega/T}+8}.
\end{eqnarray}
This implies that  the states are entangled for $\sigma<F\leq1$,
which shows that the low boundary $\sigma$ is independent of
$\alpha$ and is equal to the boundary $\tau$ in the section III.
Thus, we find an interesting result that although the parameter $\alpha$ affects the entanglement \cite{Qiyuan Pang}, it does not change the range of the parameter $F$ where the states are entangled.

\section{SUMMARY}

We have investigated how the Hawking effect and the prepared states affect the entanglement distillability of the Dirac fields
in the Schwarzschild spacetime using the Werner states
which are composed with the maximum or generically entangled states.
For the Werner states composed with the maximum entangled states, we
have found that the states are entangled for $\tau<F\leq 1$ in the
Schwarzschild spacetime. The relation $\tau=\frac{3e^{\omega/T}+5}
{6e^{\omega/T}+8}$ shows that
$\tau$ is influenced by the Hawking temperature $T$ of the black hole and
energy $\omega$ of the states. For a system with zero temperature the states
are entangled for $1/2<F\leq 1$, which is the same as that in the
inertial frame. But the states are entangled for $0.57<F\leq 1$ when
when $T\rightarrow\infty$. On the other hand, unlike in an
inertial frame, $\tau$ reduces with the increase of the energy of the
states in the Schwarzschild spacetime. For the Werner states
composed with the generically entangled states, we have shown that although
the parameter $\alpha$ affects the entanglement, it does not change
the range of the parameter $F$ where the states are entangled. The
results will help us to understand the information of black holes
more clearly.

\begin{acknowledgments}

 This work was supported by
the National Natural Science Foundation of China under Grant No
10875040; a key project of the National Natural Science Foundation
of China under Grant No 10935013; the National Basic Research of
China under Grant No. 2010CB833004, the Hunan Provincial Natural
Science Foundation of China under Grant No. 08JJ3010,  PCSIRT under
Grant No IRT0964, and the Construct Program  of the National Key
Discipline. JiLiang thanks the Kavali Institute for Theoretical Physics China for hospitality in the revised stages of this work.

\end{acknowledgments}

\end{document}